\documentclass[prl,aps,twocolumn,superscriptaddress,showpacs]{revtex4}
\usepackage{bm}
\usepackage{amssymb}
\usepackage{graphicx}

\newcommand{\two}{\frac{1}{\sqrt{2}}}

\begin{document}

\title{Experimental realization of linear-optical partial $\mathtt{SWAP}$ gates}

\author{Anton\'{\i}n \v{C}ernoch}
\affiliation{Joint Laboratory of Optics of Palack\'{y} University and
     Institute of Physics of Academy of Sciences of the Czech Republic,
     17. listopadu 50A, 779\,07 Olomouc, Czech Republic}

\author{Jan Soubusta}
\affiliation{Joint Laboratory of Optics of Palack\'{y} University and
     Institute of Physics of Academy of Sciences of the Czech Republic,
     17. listopadu 50A, 779\,07 Olomouc, Czech Republic}

\author{Lucie Bart\r{u}\v{s}kov\'{a}}
\affiliation{Department of Optics, Palack\'y University,
     17.~listopadu 50, 772\,00 Olomouc, Czech~Republic}

\author{Miloslav Du\v{s}ek}
\affiliation{Department of Optics, Palack\'y University,
     17.~listopadu 50, 772\,00 Olomouc, Czech~Republic}

\author{Jarom{\'\i}r Fiur\'{a}\v{s}ek}
\affiliation{Department of Optics, Palack\'y University,
     17.~listopadu 50, 772\,00 Olomouc, Czech~Republic}

\begin{abstract}
We present a linear-optical implementation of a class of two-qubit partial $\mathtt{SWAP}$ gates for polarization states of photons. Different gate operations, including the $\mathtt{SWAP}$ and entangling $\sqrt{\mathtt{SWAP}}$, can be obtained by changing a classical control parameter -- namely the path difference in the interferometer. Reconstruction of output states, full process tomography and evaluation of entanglement of formation prove very good performance of the gates.
\end{abstract}


\maketitle

Quantum information processing requires  precise control and manipulation of the states of quantum systems. In particular, two-qubit entangling unitary gates lie at the heart of many protocols and they are, together with single-qubit operations, sufficient for universal quantum computing \cite{Nielsen00}. The archetypal two-qubit controlled-NOT (C-NOT) gate or its equivalent has been demonstrated for several physical systems, notably trapped ions \cite{SchmidtKaler03}, nuclear magnetic spins \cite{Childs01} and polarization states of single photons \cite{Pittman03,Brien03,Gasparoni04,Zhao05,Langford05,Kiesel05,Okamoto05,Bao07}. Specific feature of the linear-optical quantum gates for photonic qubits is that the required nonlinear coupling between two photons is achieved by using interference, auxiliary photons, single photon detectors and/or conditioning \cite{Knill01,Ralph01,Pittman01,Ralph02}. A universal quantum computer can be built with the above listed resources  despite the fact that the basic gates are only probabilistic and have some finite probability of failure \cite{Knill01,Kok05}.

A major obstacle to scaling the optical schemes to higher number of qubits is the lack of a deterministic source of single photons. In the experiments, correlated photon pairs generated by means of spontaneous parametric down-conversion are utilized. Consequently, the coincidence rate decreases rapidly with the number of photons limiting current experiments to six-photon coincidences \cite{Zhang06}. This makes it currently practically impossible to implement a given desired gate by concatenating several basic quantum gates such as C-NOTs. It is therefore highly desirable to seek other means of realizing various gates with available resources \cite{Fiurasek06}.

In this Letter, we report on experimental realization of a class of entangling partial $\mathtt{SWAP}$ gates
for polarization states of photons. The gate imposes a phase shift $\phi$ to singlet Bell state  $|\Psi^{-}\rangle=\two(|H\rangle|V\rangle-|V\rangle|H\rangle)$ while it leaves unchanged the triplet Bell states $|\Psi^{+}\rangle=\two(|H\rangle|V\rangle+|V\rangle|H\rangle)$ and
$|\Phi^{\pm}\rangle=\two(|H\rangle|H\rangle\pm |V\rangle|V\rangle)$. The unitary operation can be expressed  as,
\begin{equation}
U_\phi=\Pi_{+}+e^{i\phi}\Pi_{-}, 
\label{Upartial}
\end{equation}
where $\Pi_{-}=|\Psi^{-}\rangle\langle \Psi^{-}|$ and $\Pi_{+}=I-\Pi_{-}$ are the projectors onto the antisymmetric and symmetric subspaces of the two qubits, respectively, and $I$ denotes the identity operator. $|H\rangle$ and $|V\rangle$ denote horizontal and vertical polarization state of single photon. In this notation, the diagonal and anti-diagonal linear polarization states read $|X\rangle=\two(|H\rangle+|V\rangle)$ and $|Y\rangle=\two(|H\rangle-|V\rangle)$, and the right- and left-handed circular polarizations are given by $|R\rangle=\two(|H\rangle+i|V\rangle)$,
$|L\rangle=\two(|H\rangle-i|V\rangle)$.

\begin{figure}[!b!]
\centerline{\includegraphics[width=\linewidth]{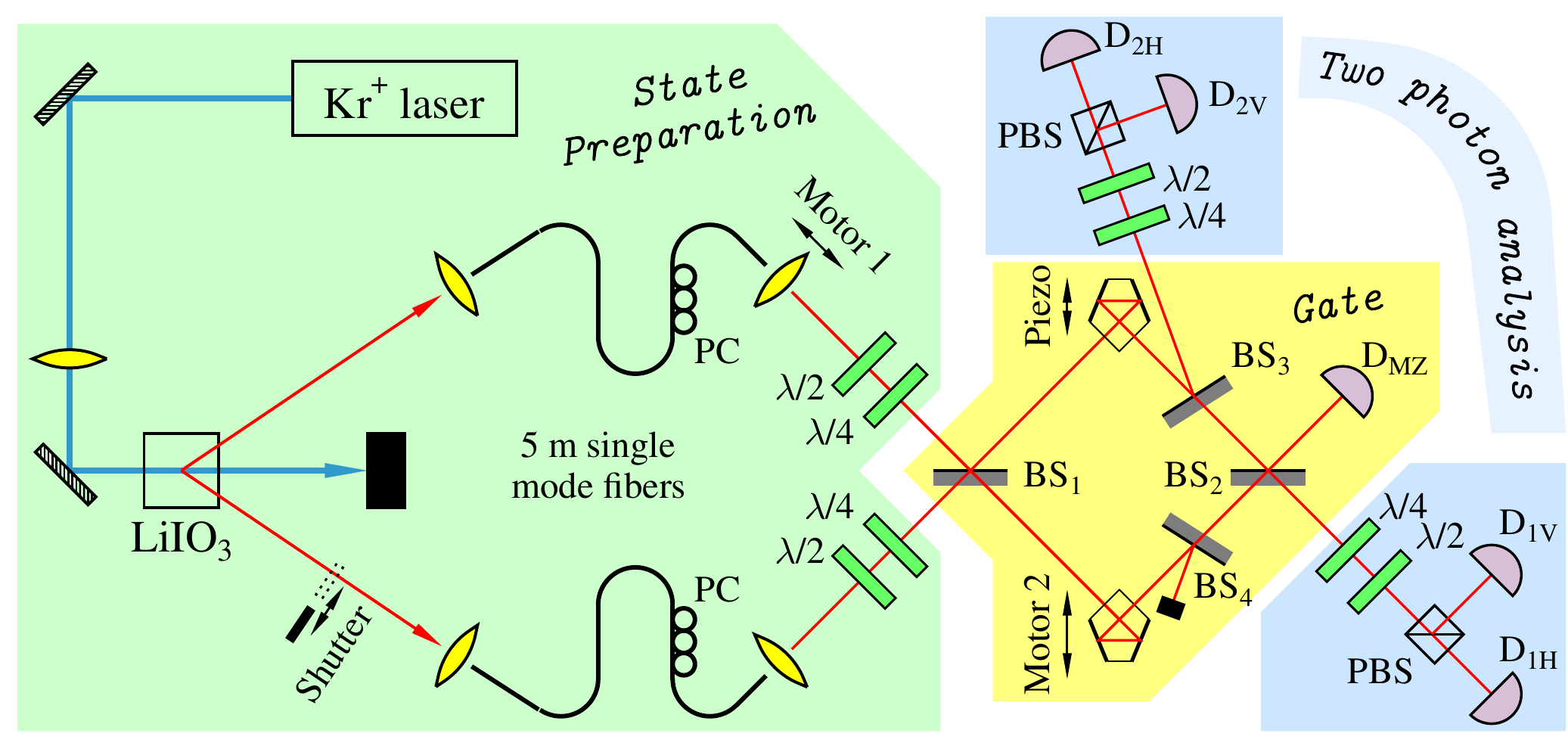}}
\caption{Experimental setup: PC - fiber polarization controller, 
        BS - non-polarizing plate beam splitter, 
        PBS - polarizing cube beam splitter, 
        $\lambda/2$ a $\lambda/4$ - wave plates, 
        D - describes a set composed of cut-off filter, collimating lens, 
            single-mode fiber and avalanche photodiode. 
\label{setup}}
\end{figure}

The gates (\ref{Upartial}) are generally locally inequivalent to the C-NOT or C-Phase gates demonstrated  previously \cite{Pittman03,Brien03,Gasparoni04,Zhao05,Langford05,Kiesel05,Okamoto05}. This means that the gate $U_\phi$ cannot be decomposed as a single C-NOT gate accompanied by single-qubit operations at the input and output. In general, sequence of up to three C-NOTs combined with single qubit transformations is required \cite{Zhang04} which would be extremely challenging to implement with present-day technology. Our approach bypasses this hurdle by combining single- and two-photon interference to directly realize the transformation (\ref{Upartial}). We utilize balanced Mach-Zehnder interferometer with two additional 50:50 beam splitters placed in each of its arm, see Fig. 1. The device operates in the coincidence basis \cite{Ralph02}, i.e. the gate is successful only if we detect a single photon in each output port, similarly as in other implementations \cite{Langford05,Kiesel05,Okamoto05}.

Let us describe the gate functioning in some detail. The photons impinge on the first beam splitter BS$_1$ where Hong-Ou-Mandel interference takes place \cite{Hong87}. 
BS$_1$ acts as a filter that distinguishes between symmetric and anti-symmetric states of the photons. If the photons are in symmetric state, than they bunch and both end up either in the upper or in the lower arm. The only way to reach the proper output ports is that they both travel through the upper arm, one photon is reflected from BS$_3$ and the other is transmitted through BS$_3$ and BS$_2$. On the other hand, if the photons are initially in the antisymmetric singlet state $|\Psi^{-}\rangle$, then each photon travels through one arm and they reach the correct output ports if one photon is reflected by BS$_3$ while the other photon is transmitted through BS$_4$ and reflected from BS$_2$. 
By changing the optical path difference the singlet state can acquire arbitrary relative phase shift $\phi$ with respect to the triplet states. It is easy to verify by direct calculation that the conditionally applied transformation is given by Eq. (\ref{Upartial}) and the probability of success is equal to $\frac{1}{8}$. By additional attenuation of the signal in the lower arm the interferometer could operate also as a partial symmetrization device conditionally applying filter $\Pi_{+}+\epsilon\Pi_{-}$ with $\epsilon <1$ \cite{Fiurasek07}. The scheme is thus very versatile and can find applications in various areas of optical quantum information processing.

The experimental setup is schematically divided into three parts, see Fig. 1. The first part is the source of the time-correlated pairs of photons. The second part represents the gate, and the last part serves for two-photon polarization analysis. Pairs of photons are generated in the process of type I spontaneous parametric down-conversion in a non-linear crystal of LiIO$_3$. The crystal is pumped by the cw Krypton-ion laser at 413~nm with power of 120~mW. The downconverted photons are coupled into two single-mode fibers and transferred to a second optical table. This is necessary in order to protect the interferometric part of the setup from vibrations caused by cooling water in the laser system. The two photons are released back to free space and the desired polarization state of each photon is set by means of half and quarter wave plates.

The second part of the setup represents the gate. As detailed above, it is formed by a standard Mach-Zehnder (MZ) interferometer with two additional balanced beam splitters in each arm. BS$_3$ splits out part of the beam for further processing, while BS$_4$ just balances the losses. Detector D$_{\rm MZ}$ monitors the interference fringes. This output is polarization independent so the signal from the detector D$_{\rm MZ}$ is used for active phase stabilization of the MZ interferometer. 

\begin{figure}
\centerline{\includegraphics[width=0.68\linewidth]{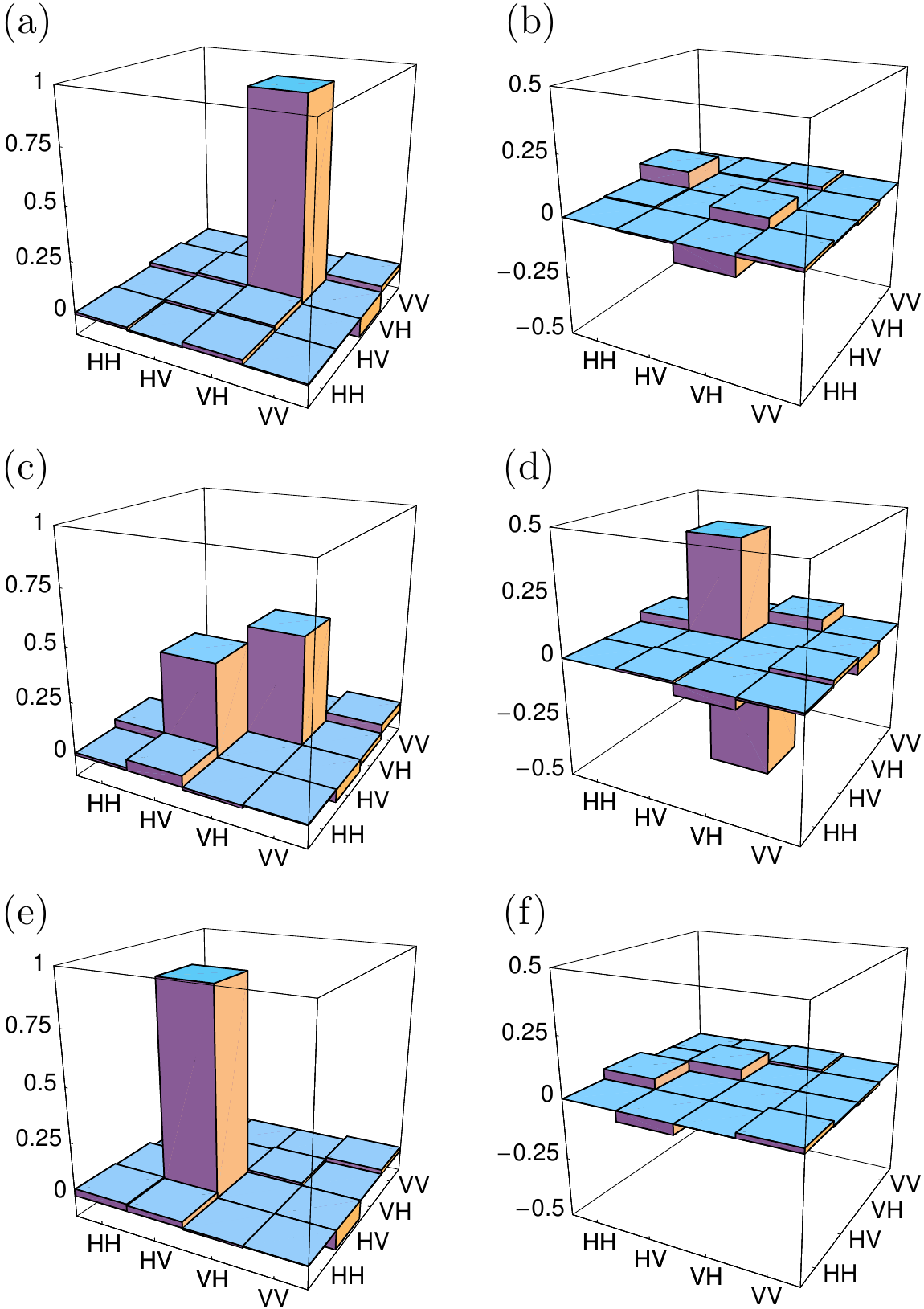}}
\caption{Real (left panels) and imaginary (right panels) parts of the reconstructed output density matrix for input state $|V\rangle|H\rangle$ are shown for three different phase shifts $\phi=0$ (a,b), $\phi=\pi/2$ (c,d) and $\phi=\pi$ (e,f).}
\end{figure}

The two-photon analysis is performed by coincidence polarization measurements between two blocks. Each block is composed of the quarter- and half-wave plate and a polarizing beam splitter (PBS), which splits the photon state into horizontal and vertical polarization components. Before detection the beams are filtered spectrally by cut-off filters at 780~nm, and geometrically by single mode fibers to ensure perfect overlap of the spatial modes. The signals from avalanche photodiodes are processed by four-input coincidence logic module.

Preliminary alignment of the setup is done in three steps. First, proper function of the photon-pair 
source is verified. A separate fiber beam splitter is used to measure two-photon
interference in Hong-Ou-Mandel (HOM) type interferometer (not shown in Fig.~\ref{setup}) 
\cite{Hong87}. Then the fiber beam splitter is replaced by two 5m long fibers. Polarization controllers (PC) 
on the fibers serve to adjust horizontal linear polarizations at the output of the fibers. 
In the second step, the beam is blocked between beam splitters BS$_3$ and BS$_2$ and the overlap of the  beams on the first beam splitter BS$_1$ is optimized. We scan the HOM interference dip
measuring coincidences between detectors D$_{1H}$ and D$_{2H}$ as a function of 
Motor 1 position. In this configuration, we have in maximum about 100 coincidence counts per second,  the visibility is about 97\%. Motor 1 position is placed to the minimum of the dip. 
As a last step we adjust the single-photon interference in the MZ interferometer. 
For this purpose, one input arm is blocked by the shutter in the source and the second beam 
splitter BS$_2$ is realigned. The lengths of the interferometer arms are balanced by 
motorized translation of one pentagon prism (Motor 2) to obtain the highest 
visibility of the interference fringes. Precise fringe phase scan is performed by piezo-driven translation  of the pentagon in the other arm and we typically observe visibility about $98\%$.

In our experiment we performed full polarization analysis of the output two-photon state for various input product states. We measured two-photon coincidence counts between detectors $\rm D_{1H} \& D_{2H}$, $\rm D_{1V} \& D_{2V}$, $\rm D_{1H} \& D_{2V}$, and $\rm D_{1V} \& D_{2H}$ for 9 (3$\times$3) combinations of the measurement basis, i.e., projections onto horizontal/vertical, diagonal/anti-diagonal and right/left circular polarizations in the two 
output arms. The unequal detector efficiencies were compensated by proper rescaling
of the measured coincidences \cite{Soubusta07}.
The state analysis was made for 36 (6$\times$6) combinations of the input polarization states $|H\rangle$, $|V\rangle$, $|X\rangle$, $|Y\rangle$, $|R\rangle$, and $|L\rangle$. We measured the gate operation for nine values of the phase shift $\phi$ in the 
MZ interferometer, $\phi=k\pi/4$, $k=0,\ldots,8$. Each of these 2916 points was measured for 
15~s. After each three measurement points the active stabilization procedure was performed as follows. The shutter blocked one arm in the source and with the help of detector D$_{\rm MZ}$ the zero phase position in the fringe pattern was updated. As a result, the interferometer was stable and the overall phase drift was negligible for a period of hours. There were only small oscillations $\lesssim 3\%$ of the period.

\begin{table}[!b!]
\begin{ruledtabular}
 \begin{tabular}{cccccc}
$\phi$ & $F_{\mathrm{av}}$ & $F_{\mathrm{min}}$ & $\mathcal{P}_{\mathrm{av}}$ & $\mathcal{P}_{\mathrm{min}}$ 

 &  $F_{\chi}$ \\
\hline
 0       & 0.960 &  0.930  & 0.957  &  0.917   &  0.946  \\
$\pi/4$  & 0.942 & 0.892  & 0.938  &  0.863   &  0.928  \\
$\pi/2$  & 0.924 & 0.876  & 0.895  &  0.804  &   0.906  \\
$3\pi/4$ & 0.929 &  0.878 & 0.908  &  0.820 &   0.914  \\
$\pi$    & 0.956 & 0.929  & 0.956  &  0.904  &   0.942  \\
$5\pi/4$ & 0.943 &  0.882 & 0.939 &  0.848 &   0.930  \\
$3\pi/2$ & 0.910 &  0.849 & 0.900 &  0.790 &   0.888  \\
$7\pi/4$ & 0.941 &  0.875 & 0.923 &  0.831 &   0.923  \\
$2\pi$   & 0.959 &  0.926 & 0.959 &  0.901 &   0.945  
 \end{tabular}
 \end{ruledtabular}
 \caption{Average and minimum fidelities and purities of the output states and the process fidelity for $9$ different phase shifts $\phi$.}
\end{table}

From the tomographically complete data we reconstructed the output two-qubit state corresponding to each input state. Standard Maximum-likelihood (ML) estimation technique was employed \cite{Jezek03}. As an example, Fig. 2 shows the reconstructed output density matrix for the input state $|V\rangle|H\rangle$ and three different values of the phase shift. For $\phi=0$, the interferometer should realize identity operation, and the output state is indeed almost identical with the input pure state $|V\rangle|H\rangle$, with fidelity $F_0=0.947$. When we set $\phi=\pi/2$, the device realizes square-root of $\mathtt{SWAP}$ ($\sqrt{\mathtt{SWAP}}$). This gate is entangling and ideally should produce maximally entangled state $\two(|V\rangle|H\rangle+i|H\rangle|V\rangle)$. The creation of this state is clearly visible in Fig. 2(c,d) with coherence appearing in the imaginary part of the density matrix. In this case the state fidelity reads $F_{\pi/2}=0.891$.  Finally, for $\phi=\pi$ we get the output of the $\mathtt{SWAP}$ gate. Again, the result agrees very well with the expected outcome $|H\rangle|V\rangle$, and $F_{\pi}=0.936$. 

A more comprehensive quantitative characterization of the gate operation is provided in Table 1,
which contains the fidelity $F_{\mathrm{av}}$ averaged over $36$ output states corresponding to input product states $|j\rangle|k\rangle$, $j,k\in\{H,V,X,Y,R,L\}$.
The table also shows the minimum fidelity $F_{\mathrm{min}}$ among these 36 states. Besides fidelity, another important characteristics of the state $\rho$ is the purity, defined as  $\mathcal{P}=\mathrm{Tr}[\rho^2]$. Ideally, all output states should be pure and $\mathcal{P}=1$. The average and minimal purities of output states are also given in Table 1. All these data confirm that the gate exhibits very good performance.
We can see that the best results are achieved for the identity and $\mathtt{SWAP}$ operations where the fidelities and purities are maximal. The most ``difficult''
operation turns out to be the $\sqrt{\mathtt{SWAP}}$, ($\phi=\pi/2$ and $\phi=3\pi/2$). But even in these cases the average state fidelity is above $90\%$.

\begin{figure}
\centerline{\includegraphics[width=0.95\linewidth]{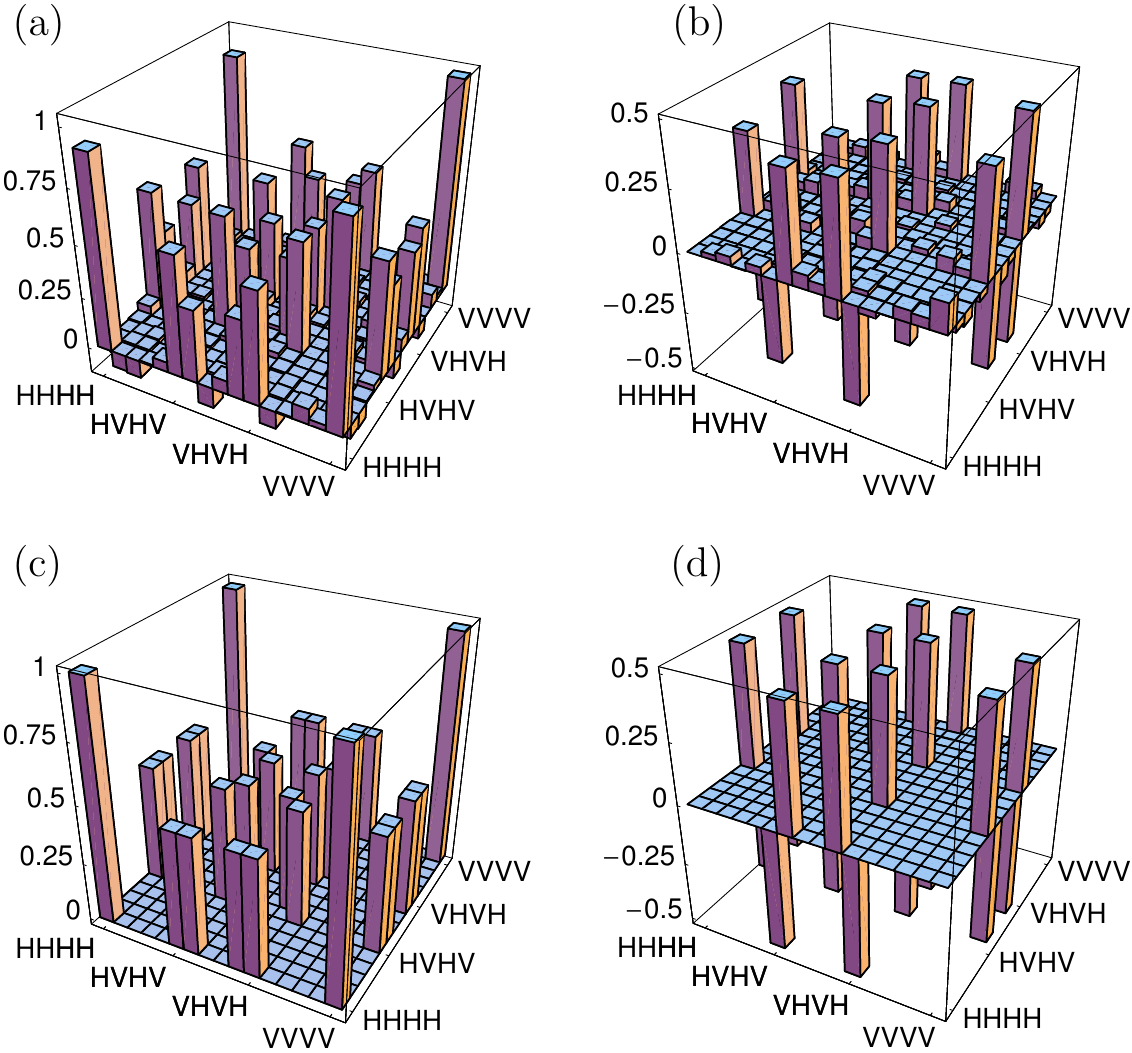}}
\caption{$\sqrt{\mathtt{SWAP}}$ gate ($\phi=\pi/2$). Real (a) and imaginary (b) parts of the reconstructed CP map $\chi$ are shown. Also the real (c) and imaginary (d) parts of the corresponding ideal map $\chi_{\mathrm{id}}$ are plotted for comparison.}
\end{figure}

The quantum gate can be fully characterized by a completely positive (CP) map. According to the Jamiolkowski-Choi isomorphism, the CP map can be represented by a positive semidefinite operator $\chi$ on the tensor product of input and output Hilbert spaces $\mathcal{H}_{\mathrm{in}}$ and $\mathcal{H}_{\mathrm{out}}$. In our case $\chi$ is thus square matrix $16\times 16$. The input state $\rho_{\mathrm{in}}$ transforms according to $\rho_{\mathrm{out}}= \mathrm{Tr}_{\mathrm{in}}[\chi \rho_{\mathrm{in}}^T\otimes I_{\mathrm{out}}]$. Since our implementation is only probabilistic, we do not assume that $\chi$ is trace-preserving but allow for general trace-decreasing map. 
Combinations of input states and corresponding measurement bases represent effective measurements performed on $\mathcal{H}_{\mathrm{in}}\otimes\mathcal{H}_{\mathrm{out}}$. Using ML estimation, we have reconstructed $\chi$ from the experimental data for $9$ different values of $\phi$. As an illustration, Fig. 3 depicts the reconstructed $\sqrt{\mathtt{SWAP}}$ gate ($\phi=\pi/2$). 
In order to quantify the quality of the operation we use process fidelity defined as
$F_\chi=\mathrm{Tr}[\chi \chi_{\mathrm{id}}]/(\mathrm{Tr}[\chi]\mathrm{Tr}[\chi_{\mathrm{id}}])$. Here $\chi_{\mathrm{id}}$
represents the ideal unitary transformation (\ref{Upartial}), which means that $\chi_{\mathrm{id}}$ is effectively a pure state. The fidelities are given in Table 1. Again, we can see that the lowest fidelity  is exhibited by the $\sqrt{\mathtt{SWAP}}$ gates, $F_{\chi}=0.906$ for $\phi=\pi/2$ and $F_{\chi}=0.888$ for $\phi=3\pi/2$. The identity and $\mathtt{SWAP}$ achieve the highest fidelities, exceeding $0.94$. 
These figures compare favorably with the values of the process fidelity of the linear optical C-NOT and C-Phase gate reported in previous experiments \cite{Langford05,Kiesel05,Okamoto05}. The unique feature of our scheme is that a whole class of inequivalent operations (\ref{Upartial}) can be reliably implemented with a single scheme, simply by changing the length of one interferometer arm with a piezo-driven translation.

\begin{figure}[!t!]
\centerline{\includegraphics[width=0.84\linewidth]{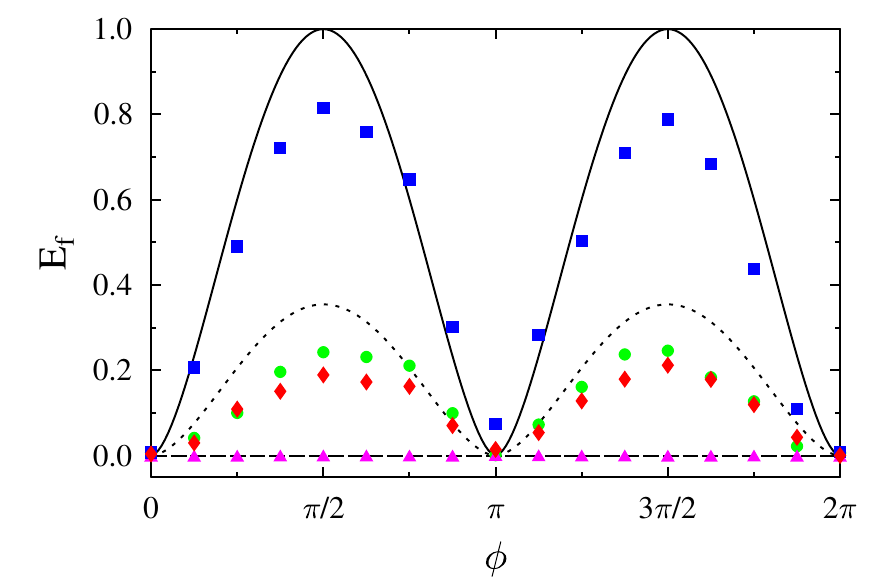}}
\caption{Dependence of the entanglement of formation $E_f$ of the output states on $\phi$ is plotted for four different input states $|X\rangle|X\rangle$ ($\blacktriangle$), $|Y\rangle|X\rangle$ ($\blacksquare$), 
$|H\rangle|X\rangle$ ($\bullet$) and $|R\rangle|X\rangle$ ($\blacklozenge$). The lines show theoretical predictions. The measurements were performed for $\phi=k\pi/8$, $k=0,\ldots,16$.}
\end{figure}

A crucial property of the entangling two-qubit gates is their ability to generate entangled states from product inputs. Consider the input product state $|\Psi_{\mathrm{in}}\rangle=|\psi\rangle(\alpha|\psi\rangle+\beta|\psi_\perp\rangle)$, where $|\psi\rangle$ is an arbitrary state and $\langle \psi|\psi_\perp\rangle=0$. The output state $|\Psi_{\mathrm{out}}\rangle=U_\phi|\Psi_{\mathrm{in}}\rangle$ reads
\[
|\Psi_{\mathrm{out}}\rangle=\alpha |\psi\psi\rangle+
\beta e^{i\phi/2}(\cos\frac{\phi}{2}|\psi\psi_\perp\rangle-i\sin\frac{\phi}{2}|\psi_\perp\psi\rangle)
\]
Its entanglement of formation is equal to the von Neumann entropy of the reduced density matrix of one of the qubits \cite{Nielsen00},  $E_f=-x\log_2(x)-(1-x)\log_2(1-x)$, where $x=\left(1+\sqrt{1-|\beta|^4\sin^2\phi}\right)/2$. Entanglement provides an additional characteristics of the gate performance to the state fidelity and purity, because states with the same purity and fidelity can generally exhibit different amount of $E_f$. In Fig.~4 we plot the entanglement of formation of the output state for four different input states $|X\rangle|X\rangle$, $|Y\rangle|X\rangle$, $|H\rangle|X\rangle$ and $|R\rangle|X\rangle$.
We observe good qualitative agreement with the theoretical expectations. The maximum entanglement $E_f=0.815$ is obtained for input state $|Y\rangle|X\rangle$ and $\phi=\pi/2$ which should ideally yield maximally entangled state in this case. The state $|X\rangle|X\rangle$ is unaffected by the transformation $U_\phi$ because it belongs to the symmetric subspace, so no entanglement is generated in this case. Finally, the inputs $|H\rangle|X\rangle$ and $|R\rangle|X\rangle$ involve two nonorthogonal states, $|\beta|^2=1/2$ and only partially entangled states are produced.

In summary, we have proposed and demonstrated linear-optical  partial $\mathtt{SWAP}$ gate for photonic qubits. The device operates with high fidelity, is easily tunable and provides a valuable addition to the toolbox of available linear-optics quantum gates.

\acknowledgments
We thank Nicolas J. Cerf for many stimulating discussions and V. Mich\'{a}lek for technical assistance. This research was supported by the projects LC06007, 1M06002 and
MSM6198959213 of MSMT.

\end{document}